\documentclass[]{emulateapj}

\usepackage[breaklinks,colorlinks,allcolors=blue]{hyperref}
\usepackage{amsmath}

\usepackage[]{units} 
\usepackage{multirow}
\usepackage{soul}
\usepackage{color}

\def\msun{{\rm\,M_\odot}}

\def\msun{{\rm\,M_\odot}}

\newcommand{\kms}{\, {\rm km\, s}^{-1}}

\newcommand{\be}{\begin{equation}}
\newcommand{\ee}{\end{equation}}

\def\h2{${\rm\,H_2}$}

\usepackage{color}

\newcommand{\EuFe}{\rm [Eu/Fe]}
\newcommand{\FeH}{\rm [Fe/H]}
\newcommand{\EuH}{\rm [Eu/H]}
\newcommand{\CFe}{\rm [C/Fe]}
\newcommand{\BaEu}{\rm [Ba/Eu]}
\newcommand{\rprocess}{\emph{r}-process }
\newcommand{\cempr}{CEMP-\emph{r} }
\newcommand{\mpr}{MP-\emph{r} }
\newcommand{\mprI}{MP-\emph{r}I }
\newcommand{\mprII}{MP-\emph{r}II }
\newcommand{\ensm}{$E_{\rm NSM}$}

\begin{document}
\title{On Neutron Star Mergers as the Source of  \emph{r}-process Enhanced Metal Poor Stars in the Milky Way}
\shorttitle{Neutron Star Mergers and \emph{r}-process Enhanced Metal Poor Stars}
\author{Mohammadtaher Safarzadeh, Richard Sarmento, and Evan Scannapieco\\
School of Earth and Space Exploration, Arizona State University}

\begin{abstract}

We model the history of Galactic \emph{r}-process enrichment using high-redshift, high-resolution zoom cosmological simulations of a Milky Way (MW) type halo. We assume that all \emph{r}-process sources are neutron star mergers (NSMs) with a power law delay time distribution.  We model the time to mix pollutants at subgrid scales, which allows us to to better compute the properties of metal poor (MP) and carbon enhanced metal poor (CEMP) stars, along with statistics of their \emph{r}-process enhanced subclasses.  Our simulations underpredict the cumulative ratios of \rprocess enhanced MP and CEMP stars (\mpr,\cempr) over MP and CEMP stars by about one order of magnitude, even when the minimum coalescence time of the double neutron stars ($t_{\rm min}$) is set to 1 Myr.  No \emph{r}-process enhanced stars form if $t_{\rm min}=100$ Myr. Our results show that even when we adopt the \rprocess yield estimates observed in GW170817, NSMs by themselves can only explain the observed frequency of \emph{r}-process enhanced stars if the birth rate of double neutron stars per unit mass of stars is boosted to $\approx10^{-4} M_\odot^{-1}$. 

\end{abstract}

\section{Introduction}

The recent aLIGO/aVirgo detection of gravitational waves from the merger of two neutron stars \citep[GW170817; ][]{Collaboration:2017kt}, and the 
subsequent kilonova observed across the entire electromagnetic spectrum \citep{Abbott:2017it,Coulter:2017ei} have confirmed that \rprocess elements are made in copious amounts in neutron star mergers \citep[NSMs;][]{Abbott:2017dq,Kasen:2017kk}. This discovery could be the sine qua non for showing that NSMs are the primary source of \rprocess elements in the Milky Way \citep{Cote:2018gj}.  

On the other hand, while it is clear that NSMs are one of the sources of \rprocess enrichment, it remains an open question if they are the most important source.   To address this question, several theoretical studies have modeled   \rprocess enrichment of a Milky Way (MW) type halo and its ultra faint dwarf (UFD) satellites by NSMs. \citet{vandeVoort:2015jw} carried out a zoom simulation of a MW type halo to $z=0$ and concluded that NSM events can explain the observed [\rprocess/Fe] abundance ratios assuming $10^{-2} M_{\odot}$ \rprocess mass is ejected into the ISM in each NSM event. \citet{Shen:2015gc} studied the sites of \rprocess production by post-processing ``Eris'' zoom simulations, and found that \rprocess elements can be incorporated into stars at very early times, a result that is insensitive to modest variations in the delay distribution and merger rates. Separately, \citet{Safarzadeh:2017dq} studied \rprocess enrichment in the context of UFDs and concluded that natal kicks can affect the \rprocess enhancement of subsequent stellar generations.

In each of these studies, it is observations of metal poor (MP) and carbon enhanced metal poor (CEMP) stars that are most constraining.  Such stars encode a wealth of information about the formation of the first stars in the universe \citep{Beers:2005kn,Frebel:2015kk}, and similarly their \rprocess enhanced subclasses (\mpr and \cempr), provide insight into the earliest \rprocess sources. Therefore, a successful theory for the source of the \rprocess should be able to explain the observed statistics of \mpr and \cempr stars in the MW's halo \citep{Barklem:2005hd,Abate:2016hb}.

In fact, the very existence of \cempr stars poses new challenges for the origin of \rprocess elements in the early universe.  These stars are believed to form at high redshifts and in low mass halos where Population III (Pop.\ III) stars have polluted the halo with their carbon rich ejecta. In such low mass halos, for a \cempr star to form, an \rprocess source that acts on a timescale similar to Pop.\ III stars (i.e., $\approx$10 Myr) is needed \citep{RamirezRuiz:2015gl}. 

Could the observed statistics of different classes of \rprocess enhanced stars be explained by NSMs as the sole source of \rprocess in the early universe?  In this study, we address this question, by carrying out a set of zoom cosmological simulations of a MW type halo and modeling NSMs as the sources of the \rprocess material. 
We improve on crucial aspects of previous such simulations on three fronts: 
(i) Modeling the coalescence timescales of  double neutron stars (DNSs) as drawn from distributions motivated by population synthesis analyses \citep{Fryer:1998bp,Dominik:2012cw,Behroozi:2014bp}. 
(ii) Identifying Pop.\ III stars by following the evolution of pristine gas in each simulation cell with a subgrid model of turbulent mixing that is crucial for properly identifying Pop.\ III stars whose ejecta are the precursor to the formation of CEMP stars \citep{Sarmento:2017du, Naiman:2018hq}; 
(iii) Adopting a high dark matter particle mass resolution in order to resolve halos where the MP and CEMP stars form in the early universe. 

The structure of this work is as follows:
In \S2 we describe our method in detail. In \S3 we present our results and compare them to observations of MW halo stars. In \S4 we discuss our results and conclusions. 
Throughout this paper, we adopt the Planck 2015 cosmological parameters \citep{Collaboration:2016bk} where $\Omega_M=0.308$, $\Omega_\Lambda=0.692$, $\Omega_b=0.048$ are total matter, vacuum, and baryonic densities, in units of
the critical density $\rho_c$, $h=0.678$ is the Hubble constant in units of  100 km/s/Mpc, $\sigma_8= 0.82$ is the variance of linear fluctuations on the
8 $h^{-1}$ Mpc scale, $n_s=0.968$ is the tilt of the primordial power spectrum, and $ Y_{\rm He}=0.24$ is the primordial helium fraction.

\section{method}

We used \textsc{ramses} \citep{Teyssier:2002fj}, a cosmological adaptive mesh refinement (AMR) code, which implements an unsplit second-order Godunov scheme for evolving the Euler equations. \textsc{ramses} variables are cell-centered and interpolated to the cell faces for flux calculations, these are then used by a Harten-Lax-van Leer-Contact Riemann solver \citep{Toro:1994gu}. 

We performed a set of zoom cosmological simulations of a MW type halo in order to address if NSMs can be considered the primary source of \rprocess enrichment in the early universe. 
We adopted three different minimum timescales for the coalescence of the DNSs: $t_{\rm min}=1$, 10, and $100$ Myrs.
We also adopted three different energy for the NS merger event and run simulations: $E_{\rm NSM}=10^{50}$, $10^{51}$, and $10^{52}$ ergs.
In all cases, we stopped the simulations at $z\approx8-9$ when reionization is complete and the formation of the metal poor stars largely diminishes.  The statistics of different classes of stars displaying a high abundance of \rprocess elements are then compared against MW's halo stars. 

\subsection{Simulation setup and Milky Way initial conditions}

To initialize our simulations, we first ran a dark matter only simulation down to redshift zero in a periodic box with a comoving size of 50 Mpc h$^{-1}$. 
Initial conditions (ICs) were generated from {\sc music} \citep{Hahn:2011gj} for a Planck 2015 cosmology.
The virial mass and radius of the halos are derived from the HOP halo finder \citep{Eisenstein:1998in}.  
We used a halo mass cut of $1-2\times10^{12}\msun$ to ensure we only identified halos with a mass similar to the MW. 
We found 275 such halos within the desired mass range in our simulation box. We further refined our MW-type halo candidates by requiring them to be isolated systems. We estimated this based on the tidal isolation parameter ($\tau_{\rm iso}$) approach \citep{Grand:2017cd}.
The isolation parameter for each halo is computed as:
\be
\tau_{{\rm iso},i} = M_{200,i}/M_{200} \times (R_{200}/r_{i})^3,
\ee
where $M_{200}$ and $R_{200}$ are the virial mass and radius of the halo of interest, and $M_{200,i}$ and $r_i$ are the virial mass of 
and distance to the $i$-th halo in the simulation, respectively. We computed $\tau_{\rm iso,max}$ for all halos with masses between $1-2\times10^{12}\msun$, 
by searching within a distance of 10 Mpc h$^{-1}$ centered on the location of each halo. 
The most isolated halos, i.e., those with lowest values of $\tau_{\rm iso,max}$ are our candidate MW-like halos. 

Next, we traced the dark matter (DM) particles within $2\times R_{200}$, for the top five candidates with the lowest values for $\tau_{\rm iso,max}$, back to the starting redshift.  
The locations of these DM particles determine the Lagrangian enclosing box. The halo with the smallest box, now our zoom region, was chosen for our simulations
to reduce the computational costs. 

For the full hydrodynamic simulations, this zoom region is refined to a base level of 12, and 13 for two different sets of simulations corresponding to a 
dark matter particle mass of $m_{\rm DM}{\approx}1.2\times 10^5\msun$ and $1.4\times10^4\msun$ respectively. The zoom region has sides  $4.4 \times 4.2 \times 6.4$ comoving Mpc h$^{-1}$ . 


\subsection{Star formation and feedback}

The stellar particle mass in the simulation is $m_*=\rho_{\rm th}\Delta x^3_{\rm min} N$ where $\Delta x_{\rm min}$ is the best resolution cell size achievable and $N$ is drawn from a 
Poisson distribution
\begin{equation}
P(N) = \frac{\bar{N}}{N!} \exp({-\bar{N}}),
\end{equation}
where 
\begin{equation}
\bar{N}=\frac{\rho \Delta x^3} {\rho_{\rm th} \Delta x^3_{\rm min}} \epsilon_*,
\end{equation}
and the star formation efficiency $\epsilon_*$ was set to 0.01 \citep{Krumholz:2007ig} in our simulations.
Setting $\rm L_{max}$, the maximum refinement in the simulation, to 24, together with $n_*=17 $ H/cm$^{3}$ as the threshold for star formation in the cells
results in a stellar particle mass of $\approx50 \msun$. This is massive enough to host the two supernovae needed to create a double neutron star.  
$\rm L_{max}$ is the maximum refinement level in the simulation. 
A further limitation on star particle formulation is that no more than 90\% of the cell's gas mass can be converted into stars.

In this study, we only modeled \rprocess elements production by NSMs and slow $s$-process channels were not modeled. Consequently, we did not model elements such as barium that have both \rprocess and $s$-process origin. Also, we did not model SN Ia because of their long average delay times of the order of 200-500 Myr \citep{Raskin:2009du}. Given the stellar particle mass ($\approx 50 M_{\odot}$), 50\% of all such particles were assumed to host one core-collapse supernova (CCSN), assigned stochastically. Therefore, half of the stellar particles generated a CCSN ejecting a total mass of $m_{\rm sn}=10 \msun$ with a kinetic energy of $E_{\rm SN}=10^{51} {\rm erg}$ 10 Myr after the star was formed. 
The metallicity yield for each CCSN is set to $\eta_{\rm SN}=0.1,$ meaning one solar mass of metals is ejected in each CCSN event. 

For each newly formed star particle, the ejected mass and energy were deposited into all cells whose centers are within 20 pc of the particle, and if the size of the cell containing the particle is greater than 20 pc, the energy and ejecta are deposed into the adjacent cells \citep{Dubois:2008iz}. Here the total mass of the ejecta is that of the stellar material plus an amount of the gas within the cell hosting the star particle (entrained gas) such that 
$m_{\rm ej} =  m_{\rm sn} + m_{\rm ent}$, and $m_{\rm ent} \equiv {\rm min}(10\, m_{\rm sn},\; 0.25\, \rho_{\rm cell}\, \Delta x^{3})$.  Similarly, the mass in metals added to the simulation is taken to be 15\% of the SN ejecta plus the metals in the entrained material, $Z_{\rm ej}\; m_{\rm ej} = m_{\rm ent}\,Z + 0.15\, m_{\rm sn}$. 

We separately tracked the metals generated by Pop III stars. These are dubbed `primordial metals' and their mass is taken to be $Z_{\rm P, ej}\; m_{\rm ej} = m_{\rm ent}\,Z_{\rm P} + 0.15\, m_{\rm sn}\, P_{\star}$ since the scalar $P_{\star}$ captures the mass fraction of the star particle that represents Pop III stars. SN feedback is the dominant driver of turbulence in our simulation and we have modeled the feedback to be purely in kinetic form. 
Lastly, we note that we do not model black hole formation and its feedback because its impact is expected to be negligible at this redshift \citep{Scannapieco:2004es,Scannapieco:2005ga,Croton:2006ew,Sijacki:2007cw}

\subsection{Cooling}
We used CLOUDY \citep{Ferland:1998ic} to model cooling at temperatures $\gtrsim 10^{4}$ K. Below this temperature we used \cite{Rosen:1995ja} and allowed the gas to cool radiatively to 100 K. 
However, adiabatic cooling can result in gas falling below this temperature.

Additionally, we supplemented the cooling in the primordial gas with an \h2 cooling model based on \cite{Martin:1996fd}. We computed the cooling rate for each simulation cell based on its density, temperature, and H$_{\rm 2}$ fraction, $f_{\rm{H_2}}$. We set the primordial \h2 fraction according to \citet{Reed:2005cd} with $f_{\rm{H_2}}=10^{-6}$. 

Although we did not explicitly model radiative transfer, we modeled the Lyman-Werner flux from our star particles since these photons destroy H$_{\rm 2}$.  We used $\eta_{\rm{LW}} = 10^{4}$ photons per stellar baryon \citep{Greif:2006bw} and assumed optically thin gas throughout the simulation volume. The total number of stellar baryons, $N_{*,b}$, was computed each step by totaling the mass in star particles assuming a near-primordial composition ($X$=0.73, $Y$=0.25). The value of $f_{\rm{H_2}}$ was then updated every simulation step:

\begin{equation}\label{eqn:fh21}
f_{H_2,new} = \frac{(f_{H_2,old} \; N_{gas} - N_{\rm LW})}{N_{gas}},
\end{equation}
where\\ 
\begin{equation}\label{eqn:fh22}
N_{\rm LW} = N_{*,b} \; \eta_{\rm LW}.
\end{equation}

We did not model the formation of H$_{\rm 2}$ since subsequent cooling is dominated by metals shortly after the first stars are formed. 
Lastly, we included a UV background  model based on \citet{Haardt:1996fq} model.

\subsection{Turbulent mixing}

We made use of the work described in \cite{Sarmento:2017du} to generate and track new metallicity-related quantities for both the gas and star particles. Specifically, for each cell in the simulation we tracked the average \textit{primordial metallicity}, $\overline Z_{\rm P}$, which tracks the mass fraction of metals generated by Pop.\ III stars, and the \textit{pristine gas mass fraction}, $P$, which models the fraction of unpolluted gas within each simulation cell with $Z<Z_{\rm crit}$.
We briefly describe these scalars here, and a more thorough discussion is presented in \cite{Sarmento:2017du}.
 
The primordial metallicity scalar, $\overline Z_{\rm P}$, tracked the metallicity arising from Pop.\ III stars. 
This scalar allowed us to track the fraction of Pop.\ III SN ejecta in subsequent stellar populations. 
Yields from Pop.\ III stars are likely to have non-solar elemental abundance ratios \citep{Heger:2002eq, Umeda:2003bh, Ishigaki:2014bk} and contribute to the unusual abundances patterns seen in the halo and UFD CEMP stars.
Knowing both $\overline Z_{\rm P}$ and the overall metallicity of the gas, $\overline Z$, allowed us to estimate the abundances of various elements, without having to track each one individually. Similarly, the elemental abundance pattern for regular metals, is accounted for by a single scalar $Z$. By tracking these values for each star particle in the simulations, and convolving them in post-processing, we can explore the composition of our star particles through cosmic time, by using a variety of yield models for both Pop.\ III and Pop. II  SNe.

Our pristine mass fraction scalar, $P$, modeled the mass-fraction of gas with $Z < Z_{\rm crit}$ in each simulation cell. Star formation took place at much smaller scales than the best resolution of typical cosmological simulations. 
Modeling $P$ allowed us to follow the process of metal mixing at subgrid scales by quantifying the amount of pristine gas within each cell as a function time.

Most simulations instantaneously update cells' average metallicity once they are contaminated with SN ejecta. However, mixing pollutants typically takes several Eddy turnover times \citep{Pan:2010db,Pan:2013hi, Ritter:2015dq}. By tracking the evolution of $P$, we can model the formation of Pop.\ III stars in areas of the simulation that would normally be considered polluted above $Z_{\rm crit}$; in effect increasing the chemical resolution of the simulation. Our model for the pristine fraction is based on accepted theoretical models \citep{Pan:2010db} and has been calibrated against numerical simulations that model the dynamical time required to mix pollutants, due to SN stirring, in an astrophysical context  \citep{Pan:2013hi}.

As stellar particles are formed within a cell, they inherit $\overline Z$, $P$ and $\overline Z_{\rm P}$, from the gas.  This allowed us to calculate the fraction of stellar mass in a given star particle that represents metal-free stars, $P_\star$, as well as the relative contributions that metals from Pop.\ III and Pop. II  stars make to the stars that are enriched, $\overline Z_{\rm P,\star} / \overline Z_{\star}.$ 

The ejecta composition for Pop. II  and Pop.\ III stars are indicated in Table~\ref{table:chem}. 
Properly accounting for turbulent mixing enables us to identify the Pop.\ III stars whose stellar yields (carbon rich ejecta) 
are different than Pop. II  stars and are responsible for the formation of CEMP stars. 
We express the abundance ratios of a star compared to that of the Sun as
\be
[A/B]=\log \left(\frac{N_{\rm A}}{N_{\rm B}} \right)_{\rm star} -\log \left(\frac{N_{\rm A}}{N_{\rm B}} \right)_{\odot},
\ee
The solar abundance of Eu ($\log \epsilon_{\rm Eu}$) is assumed to be 0.52 \citep{Asplund:2009eu} in the notation of $\log \epsilon_X=\log (N_{\rm X}/N_{\rm H})+12$ where $\rm N_{\rm X}$and $N_{\rm H}$ 
are the number densities of element X and hydrogen, respectively. Likewise for carbon we adopt $\log \epsilon_{\rm C}=8.43$ and for iron $\log \epsilon_{\rm Fe}=7.5$.

\begin{deluxetable}{ccc} 
\tabletypesize{\footnotesize} 
\tablecolumns{3} 
\setlength{\tabcolsep}{1mm}
\tablewidth{\columnwidth} 
\tablecaption{Mass fractions of metals } 
\tablehead{
\colhead{} & \colhead{${X/Z}$} & \colhead{$X/Z_P$} \\
\colhead{Element}  & \colhead{1 Gy} & \colhead{60 $M_{\odot}$ Pop.\ III SNe}} 
\startdata 
 C & 1.68 $\times 10^{-1}$ & 7.11 $\times 10^{-1}$ \\
 Fe & 5.39 $\times 10^{-2}$ & 2.64 $\times 10^{-12}$
\enddata
\label{table:chem}
\tablenotetext{}{The mass fractions of metals for selected elements used to model the normal and primordial metallicity of star particles in our simulation. 
Data for gas typical of $1\,Gyr$ post BB provided by F. X. Timmes (2016). Data for $60 M_{\odot}$ Pop.\ III SN provided by Heger (2016).}
\end{deluxetable}

We note that subgrid turbulent mixing is only modeled for the metals and not the \rprocess ejecta. However, due to the high resolution of these simulations, 
we observe a negligible difference in metal enrichment due to the computation of subgrid turbulent mixing. Therefore, we assume the same holds for 
 \rprocess material as it is treated as another scalar field similar to the metals in the code.

\subsection{Modeling neutron star mergers}\label{sec:natal_kicks}

We have modeled the formation of DNSs to take place for a tiny fraction ($10^{-3}$) of stellar particles chosen to go SNe.
This corresponds to one DNS per $10^5\msun$ of stars, that translates into a neutron star merger rate of $\approx 10^{-4}/{\rm year}$ at $z=0$ \citep{vandeVoort:2015jwa}. 

The particle chosen to host a DNS first undergoes two CCSN explosions, corresponding to the two progenitor stars. Afterwards, the particle was assigned a delay time distribution drawn from a power law $t_{\rm merge}\propto t^{-1}$  \citep[e.g.][]{Dominik:2012cw,Mennekens:2016ho} with minimum of $t_{\rm min}=$1, 10, or 100 Myr (for three separate simulations) and maximum of $t_{\rm max}=$10 Gyr respectively. 
Note that this time is \emph{after} the formation of the second neutron star in the binary. Once the merger time has elapsed we simulated the generation of \rprocess elements via a third explosion with $E_{NSM} = 10^{51}$ erg in our fiducial run, while we explored $E_{NSM} =10^{50}$, and $10^{52}$ erg cases separately. 

\subsubsection{Europium yield}

We set the fiducial value of the europium yield in the NSM events in our simulations based on the NS-NS merger detected by aLIGO/Virgo (GW170817). We adopted the estimated Eu yield of $1.5\times10^{-5}\msun$ for each NS merger event in our simulation. This number reflects the lanthanide-rich material ejected in the post-merger accretion disk outflow in a NS-NS merger event with the maximum value of 0.04 $\msun$ \citep{Cowperthwaite:2017ea} multiplied by the abundances pattern of the solar \rprocess residuals \citep{Cote:2018gj}. The disk wind ejecta could be lanthanide-rich depending on the lifetime of the hyper-massive neutron star prior to collapsing into a black hole \citep{Metzger:2014ip,Siegel:2017et}. We adopted this value since in order to answer the question of whether NSMs could by themselves explain the statistics of the \emph{r}-process
enhanced stars in the MW's halo, one needs to be conservative in the assigned yields.

\subsection{Simulation parametrizations}

We carried out five different simulations in this paper. We name the simulations as T$x$E$y$, where $x$ stands for the minimum time for coalescence of the NSMs, and $y$ stands for the 
energy for the NSM event in cgs unit. For example T10E51 stands for the simulation with minimum time for the merging of the NSMs set to 10 Myr with $E_{\rm NSM}=10^{51}$ erg . 
The dark matter particle mass resolution is $m_p\approx1.2\times10^5\msun$, and our stellar particle mass is fixed to be $50\msun$. We stopped the simulation at $z\approx8$. 
All the five simulations are summarized in Table \ref{t.sim_params}.

\begin{deluxetable}{ccccc} 
\tabletypesize{\footnotesize} 
\tablecolumns{5} 
\setlength{\tabcolsep}{1mm}
\tablewidth{\columnwidth} 
\tablecaption{Simulations parameters} 
\tablehead{
\colhead{    } & \colhead{$t_{\rm min}$(Myr)} & \colhead{$E_{\rm NSM} ({\rm erg}) $} & $z_{\rm final}$} 
\startdata 
T1E51	&1  		& $10^{51} $ 		& 8.2	\\
T10E50	&10 		& $10^{50} $		&  8.9\\	
T10E51 	&10	 	& $10^{51} $		&  8.9\\	
T10E52 	&10  		& $10^{52} $		&  8.9\\
T100E51 	&100  		& $10^{51} $		&  8.9\\
\enddata
\label{t.sim_params}
\tablenotetext{}{The characteristics of the simulations presented in this paper. We adopt the notation of T$x$E$y$ to name each simulation, where $x$ stands for the minimum time for coalescence of the NSMs, and $y$ stands for the 
energy for the NSM event in cgs unit. The simulation with minimum time for merging of 1 Myr and $E_{\rm NSM}=10^{51} {\rm erg}$ is named T1E51.
The simulation with minimum time for merging of the binaries set to 100 Myr is named T100E51.
All these three simulations have dark matter particle mass of $1.2\times10^5\msun$. The first column indicates the minimum timescale for merging of the DNSs in a power law distribution.
The second column corresponds to the energy of the NSM event, and the last column is the stopping redshift of the simulation.} 
\end{deluxetable}

\section{Results}

\begin{figure}
\vspace*{1cm}
\includegraphics[width=0.5\textwidth]{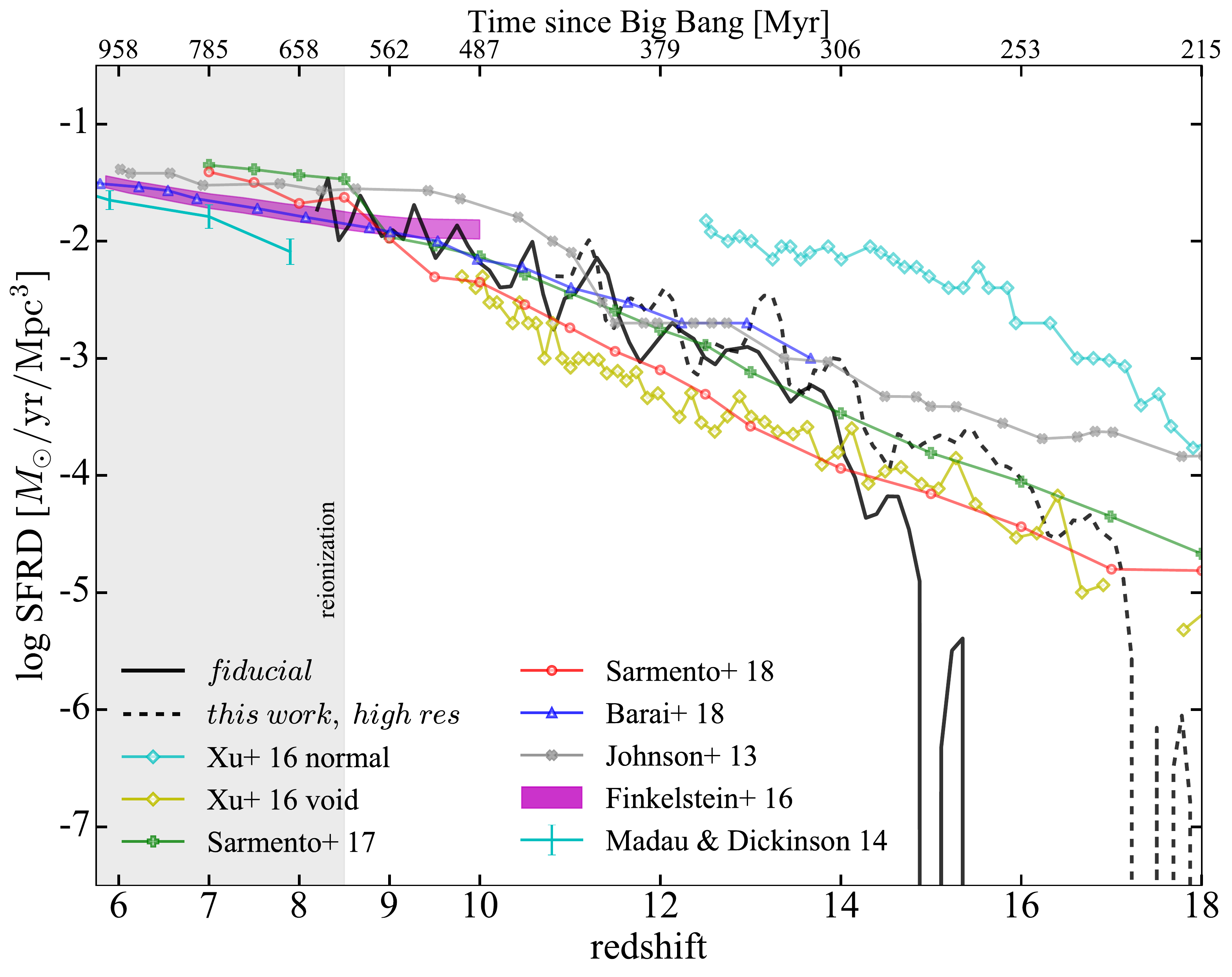}
\includegraphics[width=0.5\textwidth]{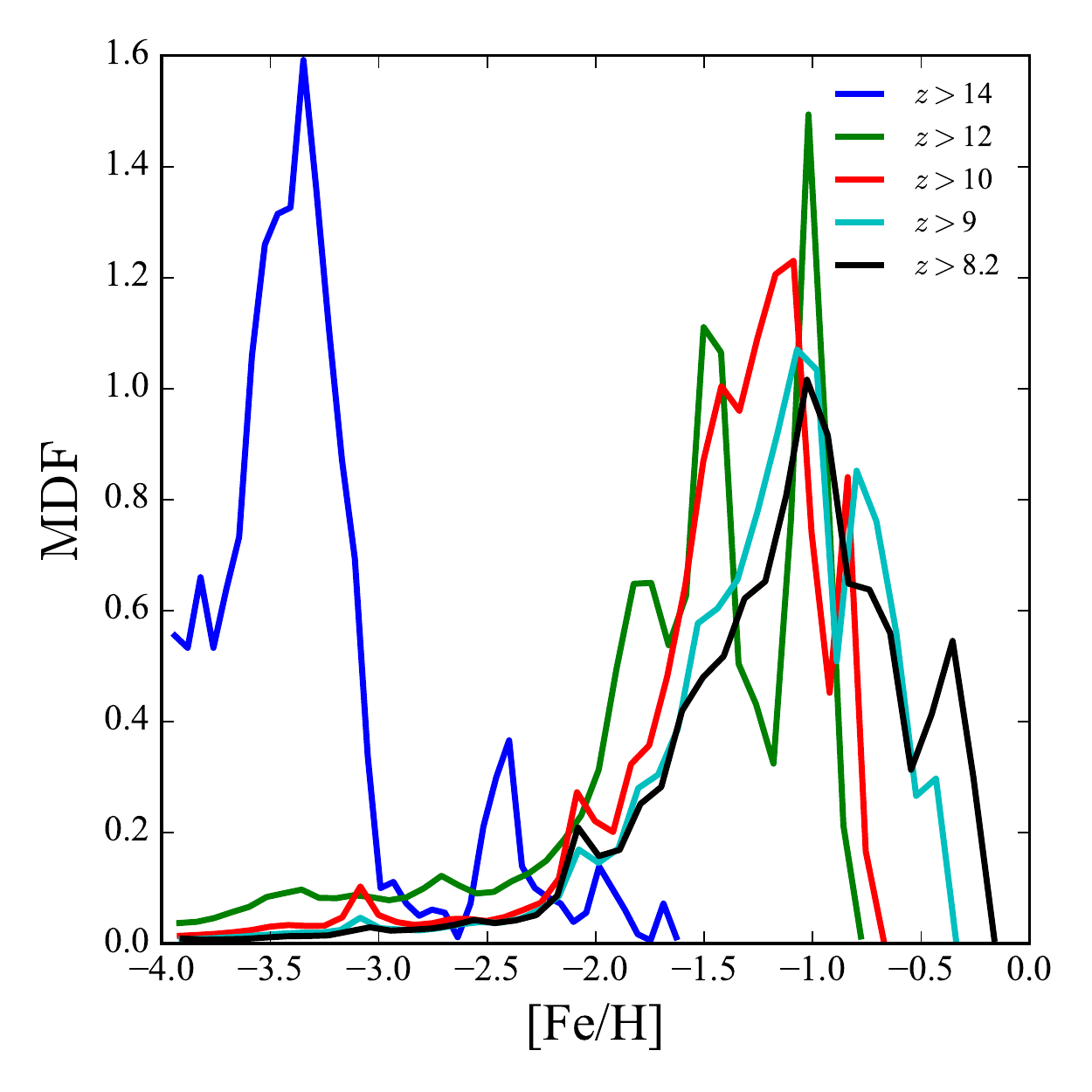}
\caption{Top panel: the comoving SFR density of the T1E51 simulation (solid black line), and the higher resolution simulation (T10E51; dashed black line ). Also shown are data from the Renaissance simulation, for both the normal and void regions \citep{Xu:2016gl}, as well as simulations by \citet{Sarmento:2017du}, \citet{Barai:2018ul}, and \citet{Johnson:2012ei}. We also include observations by \citet{Madau:2014gta} and a LF-based SFRD by \citet{Finkelstein:2016iz}. Our SFRD is in good agreement with observations and the other simulations at $z < 10$ and in reasonable agreement with the other simulations at $z > 10$ where the uncertainty is larger.
Bottom panel: the metallicity distribution function (MDF) as a function of formation redshift of the stars in the simulation: The MDF for all the stars that are
formed prior to $z$=(14,8.2) is shown in (blue, black), while other redshifts are color coded as shown in the legend.}
\label{f:fig1}
\end{figure}

We start by showing the overall star formation history of our MW type galaxy and its corresponding metallicity evolution. 
The top panel of Figure \ref{f:fig1} shows the comoving star formation rate density (SFRD) of T1E51 simulation that we ran down to redshift $z=8.2$. The cyclic SFR trend with overall increase towards lower redshift is 
characteristics of all simulations while the exact level of the SFR can vary depending on the overdensity which is re-simulated at higher resolution \citep{Xu:2016gl}. 
The improved DM mass resolution in the Renaissance simulation allows it to track star formation in smaller over densities at earlier times. Hence we see a higher SFRD at early times for the normal case as compared to simulations with lower DM mass resolution. The Renaissance simulation has a comoving resolution of 19 pc as compared to our resolution of 5 pc, however their DM particle mass is $2.9 \times 10^4$ as compared to our $1.2 \times10^5\msun$. 

The bottom panel of Figure \ref{f:fig1} shows the metallicity distribution function (MDF) for stars grouped based on their formation redshift. The MDF for stars formed at $z>14$ is shown in blue and those that are formed at $z>8.2$ shown in black.
As expected the overall metallicity increases with time while the rate of change of the MDF slows down towards lower redshifts. These are all the stars in the simulation, not categorized per halo mass.

\begin{figure*}
\vspace*{1cm}
\includegraphics[width=1\textwidth]{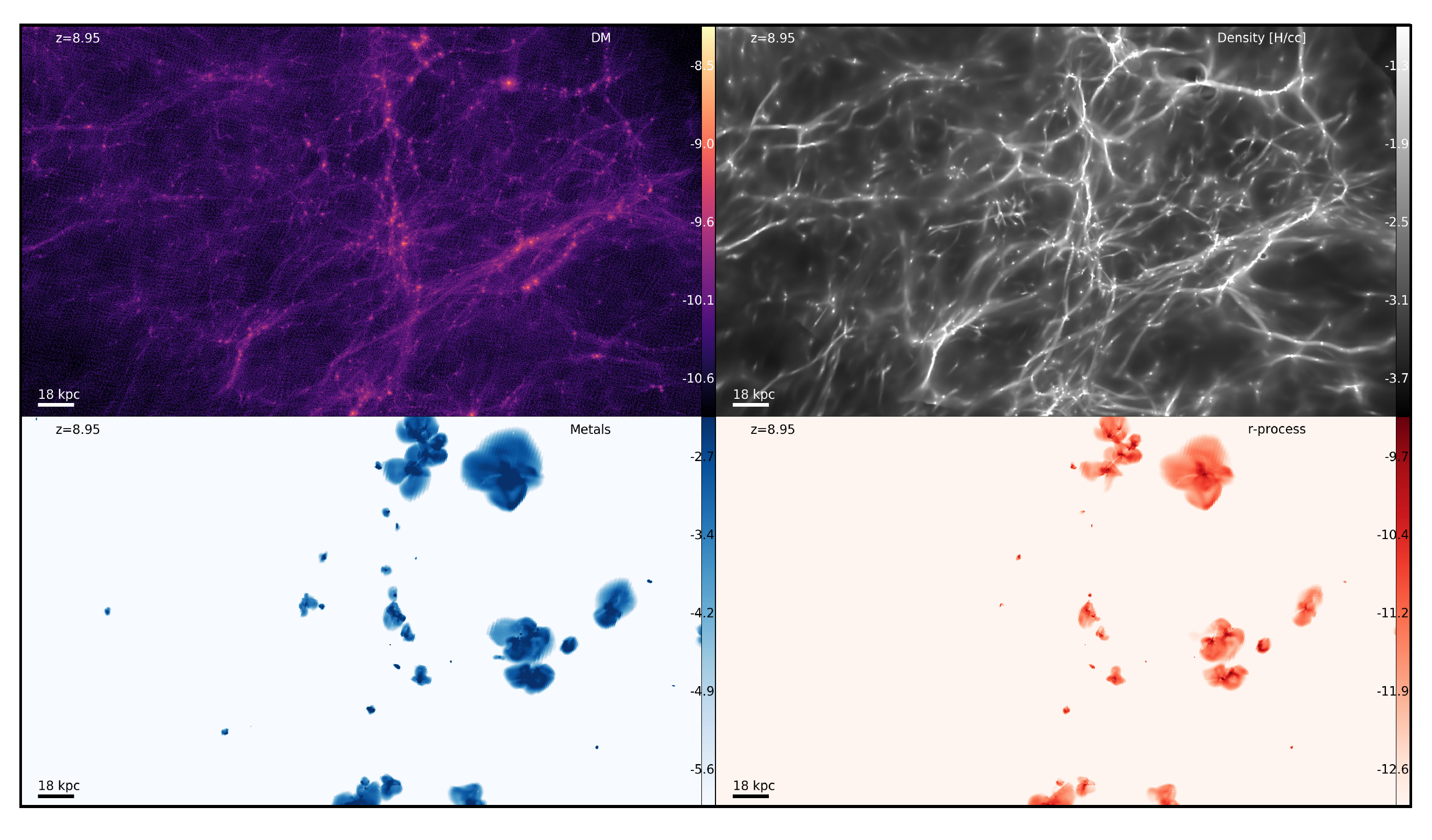}
\vspace*{-.5cm}
\caption{3D perspective snapshot of the T10E51 simulation. Clockwise from top left panel we show the dark matter, hydrogen, $r$-process, and metals distribution at $z=8.95$.
The metals are produced by CCSN and the \rprocess by the NSM, which follow a power law delay time distribution with a minimum time for merging of 10 Myr.
When comparing the metal and \rprocess distribution, we see the fact that the delay in NSMs caused some halos to be enriched with metals but no \rprocess material.}
\label{f:snapshot}
\end{figure*}

Figure \ref{f:snapshot} shows rendered images of  the dark matter, hydrogen, \rprocess, and metals in the T10E51 simulation at $z\approx9$. The fact that DNSs are born with delay time distributions causes some halos to be only enriched with metals and no \rprocess. We note that modeling DNSs' kicks will pronounce this feature 
that we present in an upcoming work.

 \subsection{Formation of CEMP stars}

Modern surveys of the Galactic halo, as well as UFDs, indicate that CEMP stars (defined as those with \CFe$>1$ and \FeH$<-1$) become more prevalent as overall metallicity decreases \citep{Beers:2005kn}. 
In fact, these surveys indicate that the fraction of CEMP stars is as high as 25\% for stars with $\FeH<-2.0$ \citep{Komiya:2007bi} and possibly as high as 40\% for stars with $\FeH<-3.5$ \citep{Lucatello:2006eg}.
\citet{Hansen:2016du} found that only about 17\% $\pm9\%$ of all the CEMP-no stars (that display no enhancement to $s$ or \emph{r}-process elements), exhibit binary orbits. Therefore, the dominant
formation scenario of the CEMP stars is not through the mass transfer from a binary companion. 
Moreover, the discovery of damped Ly-$\alpha$ systems with enhanced carbon: \citet{Cooke:2011kf,Cooke:2012dq} suggests that these stars are 
born in halos that are pre-enriched by carbon \citep{Sharma:2018km}.

\begin{figure*}
\setlength{\tabcolsep}{1mm}
\hspace{0cm}
\includegraphics[width=0.49\textwidth]{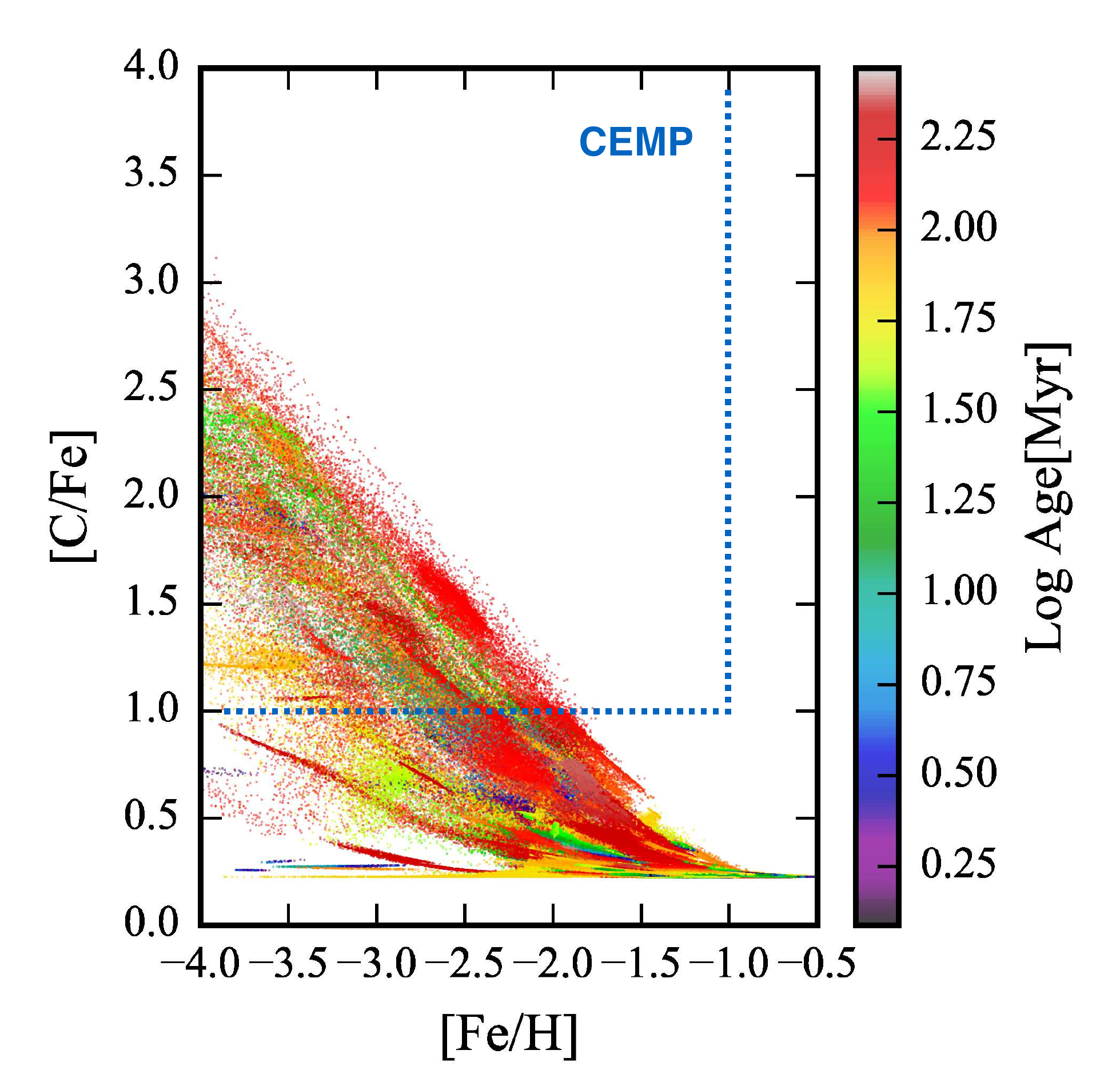}
\includegraphics[width=0.49\textwidth]{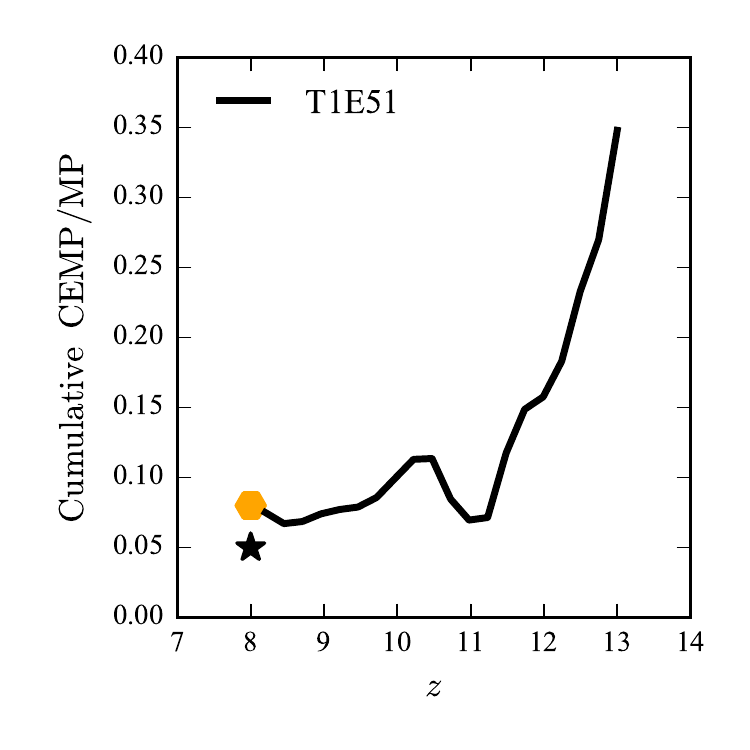}
\caption{Left panel: the distribution of the stars in $\CFe-\FeH$ plane. Each
point is a star particle color coded given its age (i.e. the red points are stars formed at the highest redshift in the simulation). The adopted Fe and C yields 
from Pop. II  and Pop.\ III  SNe is listed in Table \ref{table:chem}. Right panel: the cumulative ratio of CEMP stars to MP stars in the simulation as a function of redshift. The black line shows T1E51 simulation but the result is identical for all other simulations. The black star indicates the observed cumulative ratio of $\approx5\%$ \citep{Lee:2013by} from the SDSS/SEGUE database, and the orange hexagon is the updated analysis from \citet{Yoon:2018ej}. We note that in this comparison
we have adopted $\CFe>0.7$ for the definition for a CEMP star to be consistent with the statistics presented in \citep{Lee:2013by} and \citet{Yoon:2018ej}.}
\label{f:cemp}
\end{figure*}

Left panel of Figure \ref{f:cemp} shows the distribution of the stars in $\CFe-\FeH$ plane. Each
point is a star particle color coded given its age (i.e. the red shows the stars that formed at the highest redshift in the simulation). The adopted Fe and C yields 
from Pop. II  and Pop.\ III  SNe is listed in Table \ref{table:chem}.  Each star formation event traces a line with a negative slope in this plane. 
The oldest stars trace a line with more negative slope compared to the younger stars formed in the simulation. Since carbon is primarily generated from Pop.\ III stars,
and Pop.\ III stars are formed in metal poor regions, naturally we see higher carbon enrichment towards lower metallicities. This is consistent with the observations of the CEMP stars where higher percentage of the stars show $\CFe>1$ towards lower metallicities. The location of the stars in $\CFe-\FeH$ plane that defines a CEMP star is outlined with dashed blue line. 

Right panel of Figure \ref{f:cemp} shows the cumulative fraction of the MP stars that are CEMPs as a function of redshift.  
The black star indicates the observed cumulative ratio of $\approx5\%$ \citep{Lee:2013by} which is 
based on the SDSS/SEGUE data and consistent with other groups \citep{Frebel:2006hc,Carollo:2011ea,Placco:2014kp}. 
The orange hexagon is the updated analysis from \citet{Yoon:2018ej}. We note that in this plot we have adopted $\CFe>0.7$ for the definition for CEMP star to be consistent with the statistics presented in \citep{Lee:2013by} and \citet{Yoon:2018ej}.
The cumulative ratio of the CEMP stars to all the MP stars drops with redshift and reaches the observed ratio around $z\sim8$.

\subsection{Formation of metal poor \rprocess stars}

\begin{figure*}
\setlength{\tabcolsep}{1mm}
\hspace{0cm}
\includegraphics[width=1\textwidth]{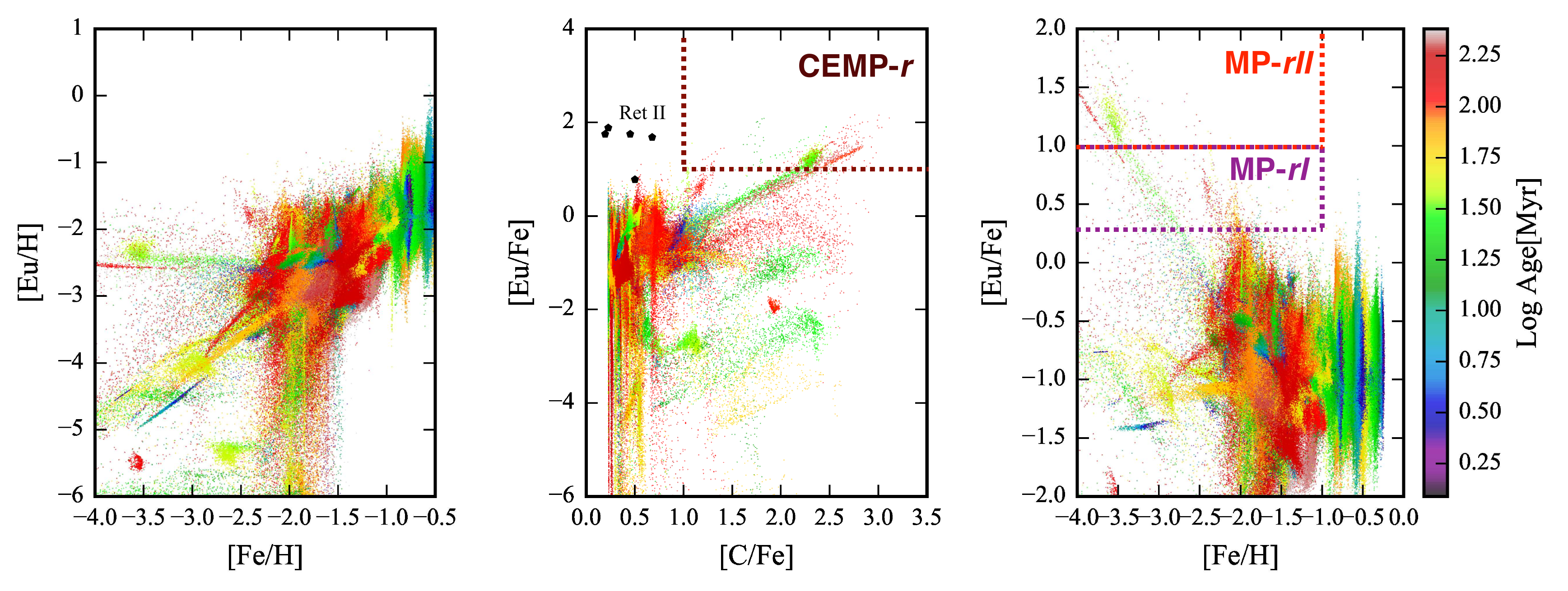}
\caption{The distribution of stars in the T10E51 simulation at $z\approx8.9$ in $\EuH-\FeH$ (left panel), $\CFe-\FeH$ (middle panel) and $\CFe-\EuFe$ plane (right panel). Each
point is a star particle color coded given its age (i.e. the red shows the stars that formed at the highest redshift in the simulation). The adopted Fe and C yields 
from Pop. II  and Pop.\ III  SNe is listed in Table \ref{table:chem}. The Eu yield per NS merger event is set to $1.5\times10^{-5}\msun$ based on the yield estimates from NS-NS merger detected by aLIGO/Virgo (GW170817). 
This reflects the lanthanide-rich material ejected in the wind ejecta from a NS-NS merger events. In the middle panel, we also show the five stars in Reticulum II whose abundances in both carbon and europium is measured.}
\label{f:fig2}
\end{figure*}

Figure \ref{f:fig2} shows the distribution of stars in in $\FeH-\EuH$, $\CFe-\EuFe$, and $\FeH-\EuFe$ planes in T10E51 at $z\approx9$.
Each data point indicates one star particle in the simulation and we show a random sample of 20\% of all the stars in the simulation box.
Each data point is color coded by the stellar age.

Close inspection of the distribution of stars in the $\FeH-\EuH$ plane shows that star formation events  trace lines with different slopes, mostly from linear in the early times to vertical at later cosmic times.
A horizontal line indicates a star formation event in a region where the gas has a dispersion in $\FeH$ but europium is well mixed, while the opposite holds for a vertical line as we go to more metal rich stars.
These are on average younger stars that have recently formed in halos
enriched by iron. These halos are enough old for the iron to be well mixed, however, a recent NSM event explains the large dispersion along the $\EuH$ axis. The correlation between $\EuH$ and $\FeH$ at
high metallicities which are shown by young stars is an imprint of the fact that stars start to form in halos where there have been pre-enrichment by both SNe II and DNSs. 

The middle panel of Figure \ref{f:fig2} shows the distribution of the stars in $\EuFe-\CFe$ plane. One can use the distribution of the stars in this plane to select \cempr stars.  
Each line traces one star formation event and as can be seen the lines have a positive slope, indicating that 
those stars that are carbon enriched, and therefore born in halos enriched with Pop.\ III ejecta, also show higher $\EuFe$ values.
This is because Pop.\ III star formation results both in supernovae that eject large amounts of carbon into their surroundings,
and DNSs that are strong sources of europium.
This leads to the observed 
correlation for old stars. As can be seen, older stars are clustered towards the lower end of $\CFe$ and do not show
the strong correlation between $\EuFe$ and $\CFe$ as is seen for the young stars. This is due to the fact that the metal production dominates over 
that of carbon in more massive halos, and in general as the formation of Pop III stars cease, the new stars in the halo are born with lower $\CFe$. In such systems, a single NSM event
will leads to large dispersion along the $\EuFe$ axis, as is observed by how the old stellar particles are clustered towards the lower end of $\CFe$.

In the middle panel, we also show the 5 stars in the ultra-faint dwarf galaxy Reticulum II \citep{Ji:2016jh} that have measured abundances in both carbon and europium. The fact that there are practically no stars in our simulation that match Ret II  abundances in both of these elements, potentially shows that the europium yield or NSM merger rate adopted as a fiducial value in our simulations needs to be boosted by a large factor. We return to this point in the next section.

The right panel of Figure \ref{f:fig2} shows the distribution of the stellar particles in the $\FeH-\EuFe$ plane. The location of the stars in this plane is used to define different category of metal 
poor \rprocess enhanced stars. 

\subsection{Comparison with observations of \rprocess enhanced metal poor stars}

Metal poor stars encode a wealth of information about the conditions in the early universe when these stars were formed \citep{Frebel:2015kk}.
Such stars are divided into two categories, \mprI and \mprII, based on the \rprocess element abundance in their spectra. 
\mprI stars are metal poor stars that show mild enhancement of \rprocess elements, namely $( 0.3<\EuFe<1 )$ and  $( \FeH < -1.5 )$.
\mprII stars are defined as those with higher levels of \rprocess abundance, namely $( 1<\EuFe )$ and $( \FeH < -1.5 )$. These two categories are outlined in the right panel of Figure \ref{f:fig2}.
Based on the  Hamburg/ESO \rprocess Enhanced Star survey \citep[HERES; ][]{Barklem:2005hd}, out of 253 metal poor stars with $-3.8<\FeH< -1.5$, 
about 5\% are \mprII and another 15-20\% are \mprI stars.
Separately, based on the SAGA database of stellar abundances, \citet{Abate:2016hb} reported that out of 451 metal poor stars with Eu and Ba abundance, 26 ($\sim6$\%) are found to belong to \mprII class.

\begin{figure*}
\setlength{\tabcolsep}{1mm}
\hspace{0cm}
\includegraphics[width=1\textwidth]{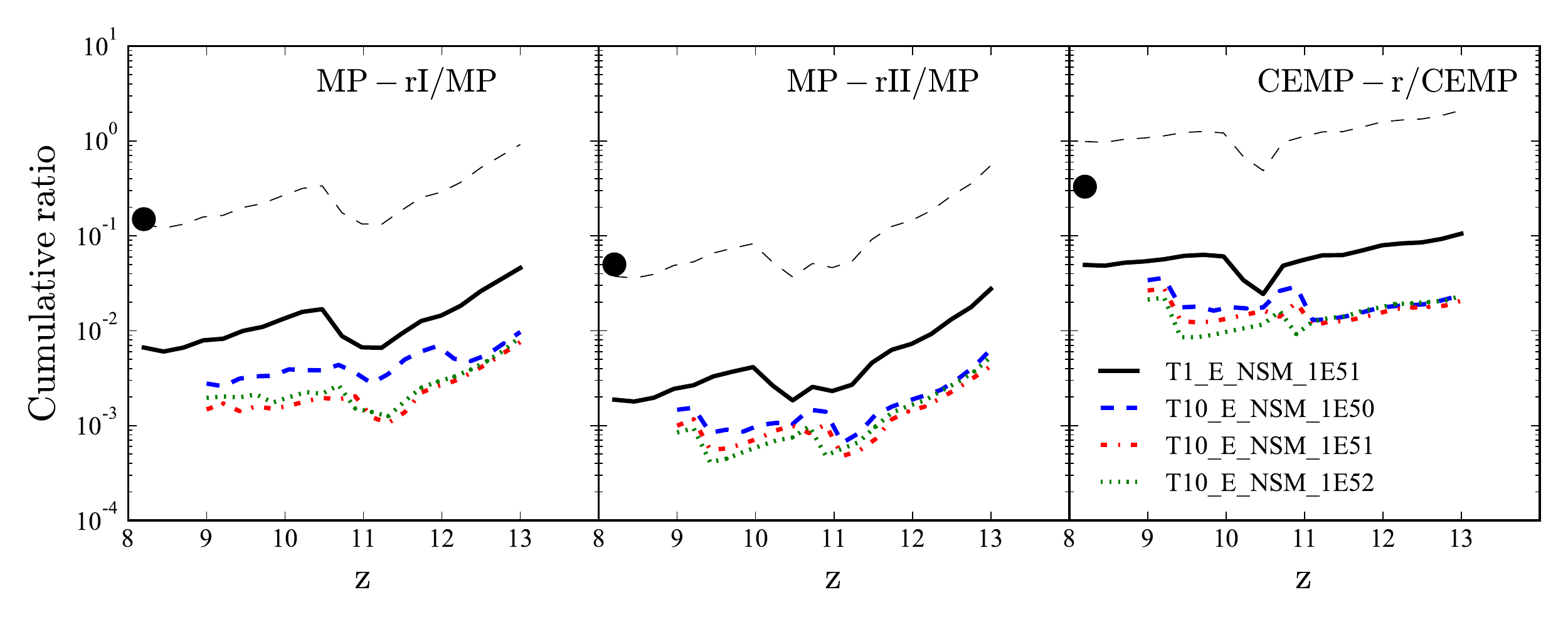}
\caption{Cumulative fraction of different class of stars in the simulation as a function of redshift. Left panel: the cumulative fraction of all the MP stars that are \mprI. Middle panel: the cumulative fraction of all the MP stars that are \mprII. Right panel: the cumulative fraction of all the CEMP stars that are \cempr. We show the results for T1E50 (solid-blue),T1E51 (dashed-green),  T1E52 (dot-dashed red), and T1E51 in solid black respectively. 
In all panels, the black dots indicate the observed ratio in the MW halo stars from \citet{Abate:2016hb}. 
The simulations severely under predict all the observed ratios by about an order of magnitude in the case of T1E51 simulation (black lines) and more so when the minimum time for merging is increased to 10 Myr.
Moreover, we see that although lower explosive energy of the NSM event helps increase the fraction of \rprocess stars, this is subdominant when compared to the impact of minimum timescale for merging.
The thin dashed lines in all three panels indicate the T1E51 result scaled by a factor of 20, translating into NSM merger rate of $\approx 2 \times10^{-4} M_\odot^{-1}$. This higher assumed higher NSM merger rate would match the observed frequency of the \rprocess MP stars but it overpredicts that of CEMP stars. }
\label{f:mp_r}
\end{figure*}

The left panel of Figure \ref{f:mp_r} shows the cumulative fraction of all the MP stars that are \mprI. This is cumulative in the sense that it indicates the fraction of all the MP stars formed by redshift $z$ that belong to the \mprI class. We show the results for T1E50 (solid-blue),T1E51 (dashed-green),  T1E52 (dot-dashed red), and T1E51 in solid black respectively. 
T100E51 simulation results in zero \mprI stars and is not shown in the plots. 
The black dot shows the ratio of \mprI over MP stars from observations of the MW's halo stars which is about 20\% \citep{Abate:2016hb}. Our simulations predict that 
the ratio is more than an order of magnitude below the level observed if the source of \rprocess is solely NSMs given the adopted rate of their formation and assigned \rprocess yield. 

Our results should be thought of in the context of the imposed delay time distributions. When a minimum timescale of 1 Myr is considered for merging of the DNSs when they are formed, given the power law distribution, the median merging timescale of the DNSs is about 100 Myr. When the minimum timescale is changed to 0 or 10 Myr, the median timescale for merging changes from 3 to 300 Myr respectively. 
These median timescales matter in that they need to be compared to a typical phase of star formation that lasts in a given MW progenitor halo. Longer merging timescales relative to star formation timescale would lead to a NSM event that does not effectively enrich the medium such that \rprocess material gets recycled into the stars formed after the event. This is either because the star formation has ceased after the NSM event,
or because a new phase of star formation occurs with a delay long enough to make the \rprocess material gets too diluted before getting recycled into the new stars.  This is clearly shown in the simulation with a  minimum merging timescale of 100 Myr, in that no \mprI stars is born in that simulation. 

The middle panel of Figure \ref{f:mp_r} shows the cumulative fraction of the MP stars that are categorized as \mprII. The lines are the same as in the left panel.  The ratio of \mprII stars to MP stars predicted in the simulation is about an order of magnitude less than the observed level in the MW halo.  

The right panel of Figure \ref{f:mp_r} shows the cumulative fraction of \cempr to all the CEMP stars. 
\cempr stars are defined as a subclass of CEMP stars with $\EuFe>1$ and $\BaEu<0$, and there are a handful of theories regarding their formation \citep{Abate:2016hb}. The location of this category of stars is outlined with dashed brown lines in the middle panel of Figure \ref{f:fig2}. Out of 56 CEMP stars with barium and europium abundances, \citet{Abate:2016hb} found 5 to be \cempr stars and 26 to be \cempr/s stars. About a few percent of all the CEMP stars are \cempr in our simulation which is an order of magnitude less than the observed frequency of this class of stars. 

The impact of the \ensm\, is understood in that lower energies tend to disperse the \rprocess material in a smaller volume and therefore the higher concentration of \rprocess leads to 
the formation of \rprocess enhanced stars. The impact of the \ensm is subdominant compared to the effect of the minimum time considered for the delay time distribution. Lower delay times (1 Myr, black line)
leads to more NSM events in a halo, while large minimum times (as in T100E51 simulation) results in formation of no \rprocess enhanced stars. 
  
In all three panels of Figure 5, the thin black dashed lines indicate an assumed NSM merger rate of $\approx 2 \times10^{-4} M_\odot^{-1}$ or equivalently a europium yield of $3\times10^{-4}\msun$ which matches the statistics of the \rprocess MP stars. This boosted NSM merger rate, however, overpredicts the same statistics for the CEMP star. The mismatch between what Eu yield is required to match the observations, either shows we need more robust statistical data for the CEMP stars, or the \rprocess MP stars have been enriched by a separate source in addition to the NSMs.
 
\section{summary and discussion}

While both core-collapse supernovae and neutron star mergers could explain the observed abundance of \rprocess elements in the Galaxy \citep{Cowan:1991ca,Woosley:1994ih,Rosswog:1999wz,Rosswog:2000nj,Argast:2004hg,Kuroda:2008dw,Wanajo:2013io,Wehmeyer:2015kl}, only \rprocess production in NSMs has been measured directly, and therefore we model the production of \rprocess through NSMs.

We performed cosmological zoom simulations of a MW type halo with dark matter particle mass resolution that can resolve 
halos of mass $\sim10^7-10^8\msun$ with spatial resolution of $\sim5$ pc. These high resolution zoom simulations are aimed
at explaining the observed high frequency of \rprocess enriched stars in the MW's halo. We assume that the only  \rprocess sources are NSMs that  are assigned delay time distribution drawn from a power law, as  predicted in population synthesis codes \citep{Dominik:2012cw}. We assign europium yield  to the NSM events representative of assuming 0.04 $\msun$ wind ejecta with solar \rprocess pattern residual possible for GW170817 \citep{Cote:2018gj}.

We track the formation of MP and CEMP stars and their \rprocess enriched counterpart \mprI, \mprII and \cempr stars and we study the impact of two parameters in our study: (i) The minimum time scale for merging after the a DNS is formed, and (ii) the impact of \ensm  on mixing the \rprocess material in a halo. Our simulations underpredict the observed ratio of \rprocess enhanced stars to their parent category by about an order of magnitude. We note that implementing the natal kicks would further reduce this enrichment level.

Our findings show that increasing the minimum timescale for merging of the DNSs results in a drop in the overall statistics of the \rprocess enhanced metal poor stars. This is due to the fact  that a longer minimum timescale for merging of the DNSs leads to lower overall NSM events during a given timespan, while increasing the median merging timescale of the DNSs.  
For example, the median timescale for merging of DNSs is (3,100, 300) Myr if the minimum timescale is set to (0, 1, 10) Myr respectively. 
Similarly, the lower the energy of the NSM event, the \rprocess material experience less mixing in the halo and this actually leads to higher levels of \rprocess enhancement for the subsequent 
stars formed in the halo. The impact of the assumed \ensm is subdominant compared to the impact that the merging timescale has on the final level of \rprocess enrichment.  

Given that with increasing the minimum time for merging from 1 Myr to 100 Myr, we are not able to form any \mprI or \cempr stars, fast merging channels for the DNSs 
seems to be a requirement to make NSMs contribute modestly to \rprocess enrichment of the Galaxy at high redshifts.

In order to match the observed enrichment, we can think of two options: (i) adopting a higher Eu yield, and (ii) increasing the DNS birth rate. 
Regarding the first option, it is highly unlikely that higher Eu yields are possible from an NSM event. The adopted yield is estimated from GW170817 \citep{Cowperthwaite:2017ea,Cote:2018gj} 
with assuming a disk ejecta of mass $0.04\msun$. However, we note that in \citet{Naiman:2018hq}, the adopted yield is three times higher that what we have adopted in our study. 

Regarding the second option, there is a large tension between the observed NS merger rates and the rates predicted from population synthesis models \citep{Belczynski:2017th,Chruslinska:2018bv}. 
The value of one merger per $10^5 M_\odot$ of stars adopted in this work corresponds to MW rate of NSMs of $\rm R_{MW}\approx10^{-4}/year$ \citep{vandeVoort:2015jwa}. This rate is on the assumption that the minimum time scale for 
merging of the binaries is 30 Myr and the final stellar mass of the MW is about $3\times10^{10}\msun$. This rate corresponds to almost the maximum rate predicted in population synthesis models with various variations, 
and about an order of magnitude above the observational estimates based on galactic double pulsars \citep{Kim:2015bi}. However, translating this rate into local rate, we would be similar to the LIGO/Virgo merger rate estimate of 
$ \rm1540^{+3200}_{-1220} Gpc^{-3} yr^{-1}$ \citep{Collaboration:2017kt}.

The NSM birth rate is subject to the details of the models implemented in the population synthesis codes 
\citep{Belczynski:2002gi,Belczynski:2008kt,Dominik:2012cw}. In the standard model assumed in these models, which mostly concerns with the assumptions governing the common envelope (CE) phase during the formation of a compact binary system, we find that with the adopted Kroupa initial mass function \citep{Kroupa:2003ig} the DNS birth rate is about 2.5 per $10^5\msun$ of stellar mass modeled. However, this birth rate can be boosted by a factor of three in variation of their standard model \citep[for example in variation 15 of][]{Dominik:2012cw}, which translate into NSM birth rates of about 6 times of what we have assumed in this study. While increasing the \rprocess yield will not impact the star formation 
history of our galaxy and simply will shift the stellar particles up and down in $\EuFe$ or $\EuH$ axis, we could not treat birth rates similar to the yields. Higher birth rates will affect the iron yield from the CCSN as their number density 
would be affected. In other words, while in our simulation there is one DNS born per 1000 CCSNe, changing that to one DNS per 100 CCSNe will significantly impact the metallicity trends in our halos. 

Based on our results higher yields or higher birth rates with fast merging timescales are needed to match the observations of the MW halo's metal poor \rprocess enhanced stars. 
Similar conclusions has been reached based on chemical evolution studies of the Galaxy \citep{Cote:2018um} and been suggested a second source of \rprocess is needed in order to explain the observed trends in MW's disk.
Moreover, the long delay between GW170817 and the star formation activity of its host galaxy, NGC 4993 \citep{Levan:2017jz}, indicates that the merger rate at short delay times is different at high redshifts.
Whether either of such choices would be consistent with the expected theoretical calculations of the \rprocess yield in NS merger events or the metallicity evolution at highest redshift remains to be explored. 
Upcoming data from the R-Process Alliance is projected to increase the detected number of  \mprII stars to 125, and over 500 new \mprI
stars in the next several years \citep{Hansen:2018ia,Sakari:2018en}. Moreover, upcoming data on the frequency of \cempr stars from high-resolution observations of a sample of approximately 200 bright CEMP stars by Rasmussen et al. (in prep) is likely to provide a much improved estimate of the frequencies of CEMP subclasses. 

\section{Future work}

We have not modeled the natal kicks of the DNSs in this work. However, their impact is expected to be significant specifically if natal kicks and delay times are \emph{not} correlated for a DNS. DNSs are thought to be the precursors of the short gamma-ray bursts (sGRBs) and the location of sGRBs with respects to galaxies in the field can provide clues into the natal kick distribution of the DNSs. 
By studying host-less GRBs, \citet{Fong:2013bc} derived natal kick velocities in the range of 20-140$\kms$ with a median value around 60$\kms$. 

From a theoretical perspective, population synthesis analysis of DNSs \citep{Fryer:1998bp} where binary systems with different initial masses for each star, initial eccentricity and orbital separation are simulated to merge and arrive at the outcome natal kick velocity after the second star goes off as a SNe. Such models arrive at natal kick distributions with an exponential profile and a median of $180 \kms$ \citep{Behroozi:2014bp}. \citet{Safarzadeh:2017dw} studied the impact of DNS's natal kick on the Galactic \rprocess enrichment and concluded that almost 50\% of all the NSMs that have occurred in the star formation history of a MW type system do not contribute to the \rprocess enrichment as the DNSs merge well outside the galaxy's effective radius. 

For systems with shallow potential wells such as the Ultra Faint Dwarfs \citep[UFDs, with halo mass of $\sim10^{7-9}\msun$; ][]{Simon:2011dm}, and their progenitors at high redshifts \citep{Safarzadeh:2018fi} small natal kicks on the order of 10-20 $\kms$ can make DNS escape their hosts \citep{Kelley:2010iq,Safarzadeh:2017dw}. This can severely impact the level of enrichment of the halos and should leave a clear mark on \cempr/CEMP ratio specifically since CEMP stars only formed early on before the halo is heavily enriched with metals, and would be almost impossible to make \cempr stars if the DNSs escape their host halo. 

Another avenue to improve on the present work would be to model the $s$-process enrichment of the stars so that comparisons could be made with the statistics of the CEMP-$s$ stars in the MW.  For that we would need to model the formation of the AGB stars \citep{Sharma:2018km}.
This work could be expanded to a whole suite of MW type halos in large simulations such as Auriga \citep{Grand:2017cd} and Caterpillar suite of simulations \citep{Griffen:2016kn} to achieve a reliable halo-to-halo scatter.

\section{Acknowledgements}
We are thankful to the referee for useful comments. We are also thankful to Enrico Ramirez-Ruiz, Tim Beers, Brian Metzger, Jeff Andrews, Daniel Siegel, and Tassos Fragos for valuable discussions. 
This work was supported in by the National Science Foundation under Grants AST-1715876 \& PHY-1430152 (the Joint Institute for Nuclear Astrophysics - Center for the Evolution of the Elements), and NASA theory grant NNX15AK82G. 
We thank the Texas Advanced Computing Center (TACC) and the Extreme Science and Engineering Discovery Environment (XSEDE) for providing HPC resources.
\bibliographystyle{mnras}
\bibliography{biblio,the_entire_lib}

\end{document}